\documentclass[12pt,a4paper]{article}
\usepackage{fullpage}
\usepackage{multirow}
\usepackage{graphicx}
\usepackage{amsmath}
\usepackage{slashed}
\newcommand{\helac}{\texttt{HELAC}}
\newcommand{\helacdipoles}{\texttt{HELAC-Dipoles}}
\newcommand{\helaconeloop}{\texttt{HELAC-Oneloop}}
\newcommand{\helacphegas}{\texttt{HELAC-PHEGAS}}

\newcommand{\amcatnlo}{\texttt{aMC@NLO}}
\newcommand{\pythia}{\texttt{PYTHIA}}
\newcommand{\herwig}{\texttt{HERWIG}}
\newcommand{\lhapdf}{\texttt{LHAPDF}}

\newcommand{\madloop}{\texttt{MADLOOP}}
\newcommand{\powheghelac}{\texttt{POWHEG-HELAC}}
\newcommand{\powhegbox}{\texttt{POWHEG-Box}}
\newcommand{\fastjet}{\texttt{FastJet}}

\newcommand{\mev}{\ensuremath{\,\mathrm{MeV}}}
\newcommand{\gev}{\ensuremath{\,\mathrm{GeV}}}
\newcommand{\tev}{\ensuremath{\,\mathrm{TeV}}}

\newcommand{\pTl}{\ensuremath{p_{\perp}^{\ell}}}

\newcommand{\pTj}{\ensuremath{p_{\perp,j_1}}}

\newcommand{\pTmiss}{\ensuremath{p_{\perp}^{\rm miss}}}
\newcommand{\HT}{\ensuremath{H_{\perp}}}

\newcommand{\bt}{\ensuremath{\bar{{\rm t}}}}
\newcommand{\bb}{\ensuremath{\bar{{\rm b}}}}
\newcommand{\bq}{\ensuremath{\bar{{\rm q}}}}

\newcommand{\NLO}{{\rm NLO}}
\newcommand\fig[1]     {Fig.\,{\ref{#1}}}

\newcommand\tab[1]     {Table~\ref{#1}}

\begin{document}
\begin{titlepage}

\begin{center}
\large\bfseries
Standard Model Higgs boson production in association with a
top anti-top pair at NLO with parton showering
\\[0.6cm]
\normalsize\normalfont
M.~V.~Garzelli$^1$, A.~Kardos$^1$, C.~G. ~Papadopoulos$^2$ and
Z.~Tr\'ocs\'anyi$^{1,3}$
\\[0.4cm]
\small\itshape
$^1$ Institute of Physics, University of Debrecen, H-4010 Debrecen P.O.Box 105, Hungary \\[0.15cm]
$^2$ NCSR Demokritos, Institute of Nuclear Physics, GR-15310 Athens, 
Greece \\[0.15cm]
$^3$ Institute of Nuclear Research of the Hungarian Academy of Sciences, Hungary
\\[1.2cm]
\small\itshape
\end{center}

\begin{abstract}
We present predictions for the production cross section of a Standard
Model Higgs boson in association with a t\bt\ pair at next-to-leading
order accuracy using matrix elements obtained from the \helaconeloop\
package. The NLO prediction was interfaced to the \pythia\ and \herwig\
shower Monte Carlo programs with the help of \powhegbox,
allowing for decays of massive particles, showering and
hadronization, thus leading to final results at the hadron level. 
\\[0.6cm]
PACS {\bf 12.38.-t} -- {Quantum ChromoDynamics}\\
PACS {\bf 13.87.-a} -- {Jets in large-$Q^2$ scatterings}\\
PACS {\bf 14.65.Ha} -- {Top quarks}\\
PACS {\bf 14.80.Bn} -- {Standard Model Higgs boson}\\
\end{abstract}

\nonumber

\end{titlepage}

\section{\label{sec:introduction} Introduction}

In recent years a plethora of NLO QCD calculations have been
presented in the literature. High-energy experiments, notably at the LHC,
have and will be benefited by the progress in our computational ability
to deal with higher order corrections to scattering amplitudes with many
partons involved. In order though to get the optimum benefit and to
produce predictions that can be directly compared to experimental data at
the hadron level, a matching with parton showers and hadronization
is ultimately inevitable. 

According to the literature, two methods exist that deal with the matching
of parton showers (PS) to matrix element (ME) calculations at NLO accuracy,
MC@NLO~\cite{Frixione:2002ik, Frixione:2010wd} 
and POWHEG \cite{Nason:2004rx, Frixione:2007vw}. 
At the same time, several 
computational frameworks for NLO QCD calculations have emerged. 
In constructing a general interface of 
PS to ME computations with NLO accuracy, we
have chosen to combine the POWHEG and HELAC-NLO~\cite{Bevilacqua:2010mx} 
approach, respectively. The first application of our project was presented in
Ref.~\cite{Kardos:2011qa}.

In this letter we continue our investigation and validation of such an
interface by studying Higgs boson production in association with
a t\bt\ pair. There are several reasons to perform this work.
On the phenomenological side, there is still interest to use this
channel to search for Higgs boson in the low mass 
regime~\cite{Dittmaier:2011ti}. On
the theoretical side, the appearance of recently developed tools as
implemented in \amcatnlo\ \cite{Frederix:2011zi}, 
also used in the analysis of Higgs
boson production in association with a t\bt\ pair, offers a unique
opportunity for cross-checking and validating our computations and
the corresponding computer programs.

\section{\label{sec:method} Method}

We performed our calculations using the \powhegbox\ \cite{Alioli:2010xd},
a flexible computer framework implementing the POWHEG method, already
used in the study of different processes also by other authors
(among the others, the most recent applications include  
Ref.~\cite{Alioli:2010xa, Melia:2011gk, Oleari:2011ey, Melia:2011tj}).
In this work the following ingredients were requested and provided
to \powhegbox\ as input:
\begin{itemize}
\itemsep -2pt
\item
The flavor structures of the Born (gg$\to$ Ht\bt, q\bq $\to$ Ht\bt, \bq
q$\to$ Ht\bt)
and real radiation emission
(q\bq$\to$ Ht\bt g, gg$\to$ Ht\bt g, \bq g$\to$ Ht\bt\bq,
g\bq$\to$ Ht\bt\bq, \bq q$\to$ Ht\bt g, qg$\to$ Ht\bt q, gq$\to$ Ht\bt q)
subprocesses (q$\in$\{u,d,c,s,b\}) were specified.
\item
The Born-level phase space was obtained by using the invariant mass of the
t\bt\ quark pair and four angles. 
\item
Squared ME's with all incoming momenta for the Born and the
real-emission processes were built using amplitudes computed by
 \helaconeloop~\cite{vanHameren:2009dr}, 
on the basis of skeleton structures generated by this same program 
and \helacphegas\ \cite{Cafarella:2007pc}, respectively.
The ME's in the
phy\-si\-cal channels were obtained by crossing.
\item
Color structures for building the color-correlated Born 
amplitudes were taken from \helacdipoles~\cite{Czakon:2009ss}.
\item
Spin-correlated Born amplitudes were projected from the helicity basis 
to the Lorentz one by using the polarization
vectors.  
\item
\helaconeloop\ was also used for computing the finite part of the
virtual correction contributions in dimensional regularization,
on the basis of the OPP method~\cite{Ossola:2006us} 
complemented by Feynman Rules for
the computation of the QCD $R_2$ Rational terms~\cite{Draggiotis:2009yb}.
\end{itemize}

With this input \powhegbox\ was used to generate events at the Born
level plus first radiation, stored in
Les Houches Event Files (LHEF)~\cite{Alwall:2006yp}. 
Then, one can choose any shower
Monte Carlo (SMC) program for generating events with hadrons, just by
showering the previous ones and applying hadronization and decay models
(it is worth mentioning that 
\powhegbox\ was written in the effort of being shower detail independent).     
We worked with both the \pythia\ \cite{Sjostrand:2000wi,Sjostrand:2006za}
 and \herwig\ \cite{Corcella:2000bw,Corcella:2002jc} SMC generators. 

In the \powhegbox\ the first emission is the hardest one measured by 
transverse momentum which is also the ordering variable in \pythia.  
However, if the ordering variable in the shower is different from 
this one, 
such as in \herwig, 
the hardest emission can be different from the first one. 
In such cases, in order to face this issue, \herwig\ is constrained to 
discard shower evolutions (vetoed shower) with larger transverse
momentum in all splittings occuring after the first emission, that
is computed by \powhegbox\ . In
addition, a truncated shower, simulating wide-angle soft emission before
the hardest emission, is in principle also needed 
for POWHEG matching~\cite{Nason:2004rx}, 
but its effect turned out to be small~\cite{LatundeDada:2006gx}. 
As there is no implementation of truncated shower in \herwig\ 
using external LHEF, 
the effect of the truncated shower is absent from our predictions.

\section{\label{sec:checks} Checks}

The Born and the real ME's were tested 
in randomly chosen phase-space points against 
results from 
\helacphegas\ . 
The consistency among real emission, Born, color-correlated and
spin-correlated ME's was tested in 
randomly chosen real-emission
phase-space regions, 
by examining the ratio between the real-emission amplitudes 
in soft and collinear limits and the corresponding subtraction terms, 
automatically computed by \powhegbox\ as approaching 
the related singularities.

The process presented here was studied extensively at the NLO accuracy in
the literature~\cite{Beenakker:2001rj,Beenakker:2002nc,Reina:2001sf,
Reina:2001bc,Dawson:2003zu}, which enabled us to make detailed checks
of our calculation.  

\begin{table}[h!]
\begin{center}
\begin{tabular}{|c|c|c|c|}
\hline
\hline
 & $m_H$, GeV & $\sigma^{\mathrm{NLO}}_{\mathrm{lit.}}$, fb & $\sigma^{\mathrm{NLO}}_{\mathrm{PH}}$, fb \\
\hline
\hline
\multirow{2}{*}{TeVatron} & $120$ & $4.857(8)$ & $4.851(3)$ \\
                          & $140$ & $2.925(4)$ & $2.918(2)$ \\
\multirow{2}{*}{$(\sqrt{s} = 2$\,TeV)} & $160$ & $1.806(2)$ & $1.797(2)$ \\
                          & $180$ & $1.132(1)$ & $1.128(1)$ \\
\hline
\multirow{2}{*}{LHC}      & $120$ & $701.5(18)$ & $701.3(8)$ \\
                          & $140$ & $452.3(12)$ & $452.8(5)$ \\
\multirow{2}{*}{$(\sqrt{s} = 14$\,TeV)}& $160$ & $305.6(8)$ & $305.0(4)$ \\
                          & $180$ & $214.0(6)$ & $213.9(3)$ \\
\hline
\end{tabular}
\end{center}
\caption{\label{tab:nlocomp1ttH}Comparison of the \NLO\ cross sections by \powheghelac\ (PH)
to the predictions of Ref.~\cite{Beenakker:2002nc} (lit) at the default scale
$\mu_R = \mu_F = \mu_0$.}
\end{table}
\begin{figure*}[t!]
\begin{center}
\includegraphics[width=0.75\linewidth]{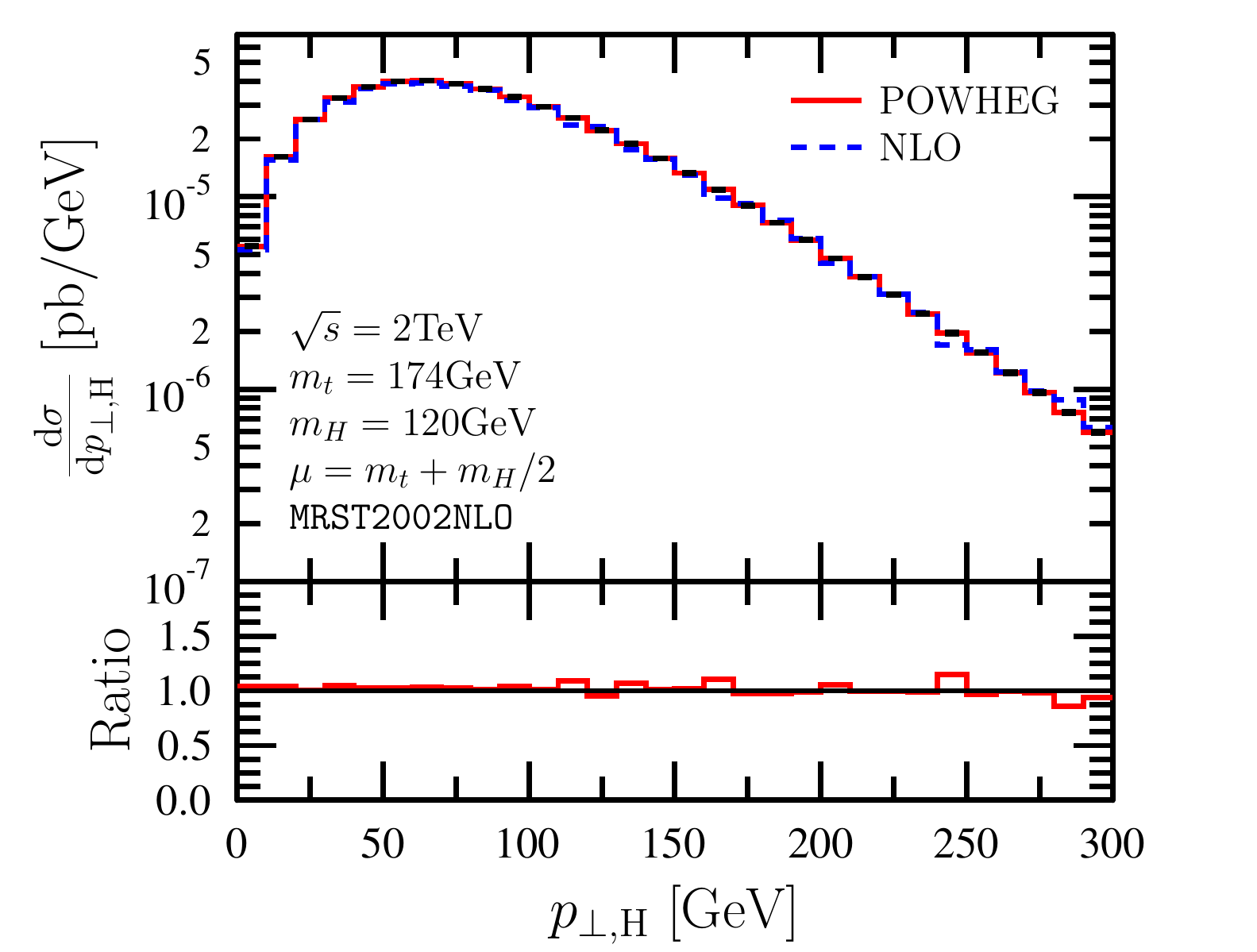}
\includegraphics[width=0.75\linewidth]{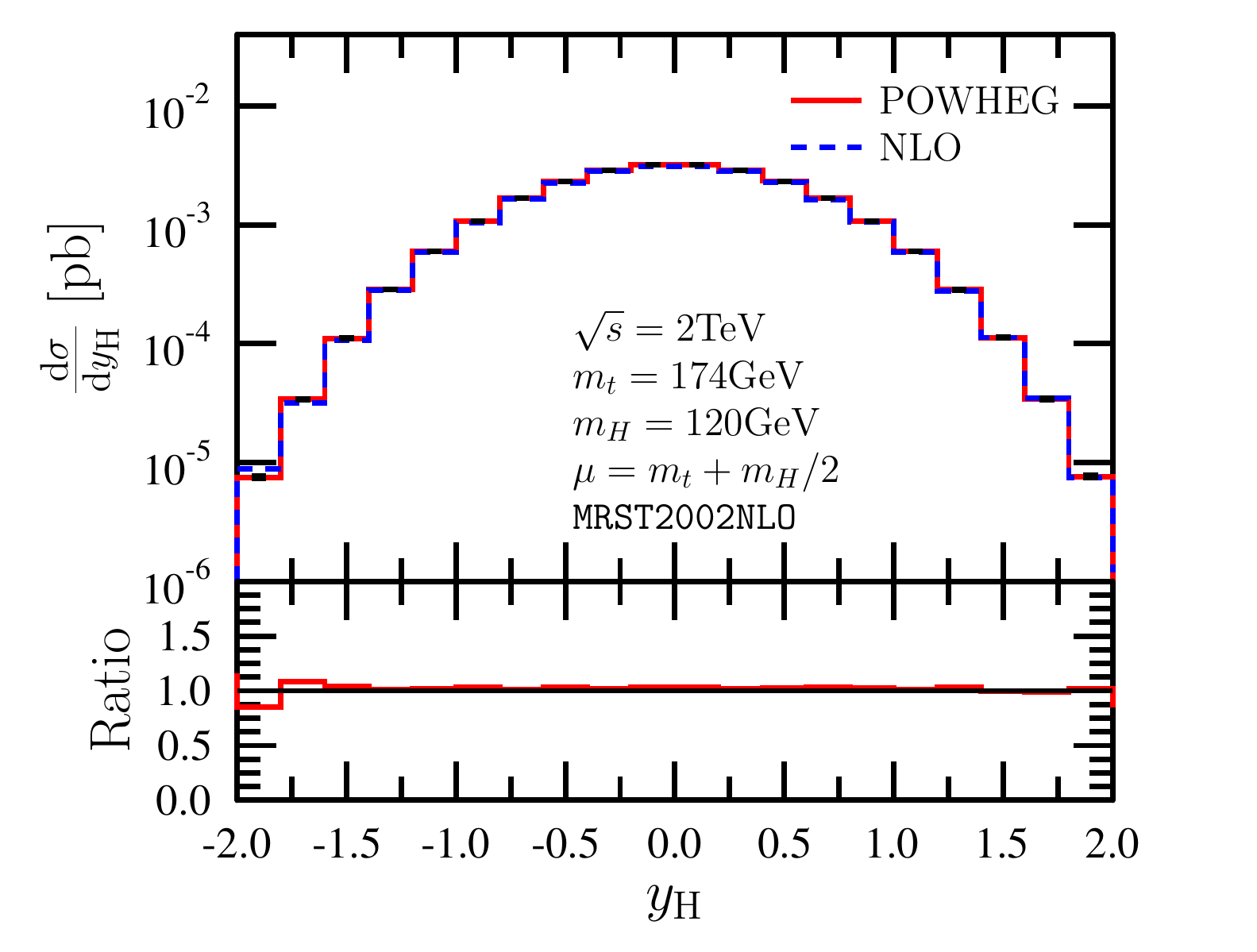}
\caption{Transverse momentum and rapidity distributions of the Higgs
boson at the Tevatron from our calculation by means of \powheghelac\
(POWHEG) and from Ref.~\cite{Beenakker:2002nc} (pure NLO).
The lower panels show the ratio of the results (POWHEG/NLO).}
\label{fig:distcomp_ttH}
\end{center}
\end{figure*}

As for the virtual contribution, we compared our amplitudes to the
predictions of \madloop\ using the input parameters and phase space
point specified in Ref.~\cite{Hirschi:2011pa} for different renormalization
scale choices \cite{Rikkert:private}, and in all cases we found 
agreement up to 5--6 digits.  

We then compared the cross-sections at the \NLO\
accuracy to the predictions of Ref.~\cite{Beenakker:2002nc}. 
At this purpose, we used the \texttt{MRST2002nlo}~\cite{Martin:2002aw} 
PDF set, provided by \lhapdf\ \cite{Whalley:2005nh}, 
with a 2-loop running $\alpha_s$, 5 light
flavours and $\Lambda_5^{\overline{\mathrm{MS}}} = 239\mev$, and the following
parameters: $m_H = 120\gev$, $m_t=174\gev$,
$G_F=1.16639\cdot 10^{-5}\gev^{-2}$. The renormalization and
factorization scales were chosen equal to $\mu$, where $\mu$ was varied
in the $\mu_0/4$ -- $4\mu_0$ interval, with default scale being
$\mu_0 = m_t + m_H/2$.  The calculations were performed at the planned
LHC energy of $\sqrt{s} = 14\tev$.  The results of this comparison are
presented in \tab{tab:nlocomp1ttH}, showing agreement.  We also made a
comparison to the predictions of Ref.~\cite{Dawson:2003zu}, using the
\texttt{CTEQ5M} PDF set from \lhapdf, a 2-loop running $\alpha_s$
with $\alpha_s(M_Z) = 0.118$, $m_t$ and $G_F$ as given above, and 
again found agreement within their quoted uncertainty.

Finally, we also compared differential distributions from events
already including first radiation emission with the \NLO\ predictions 
of Ref~\cite{Beenakker:2002nc}. We found agreement for all
distributions presented in that paper. As examples, we show 
in \fig{fig:distcomp_ttH} the transverse momentum and rapidity 
distributions of the Higgs boson in the Tevatron setup~\footnote{ 
Note that these comparisons refer to
distributions without normalization, as opposed to the plots in the
original publication.}.

\section{\label{sec:showereffect} The effect of the shower}

In interfacing NLO calculations to SMC programs we mainly aim
at estimating the effects of showers and hadronization, therefore, we
analyzed the events at two different stages of evolution:
\begin{itemize}
\item[]
{\bf Decay:} starting from the events collected in LHEF produced
by \powheghelac, we just included on-shell decays of t-quarks 
and the Higgs boson, as implemented in \pythia, and further decays
of their decay products, like charged leptons (the $\tau$ is
assumed to be unstable), $W$ and $Z$, turning off any shower and 
hadronization effect.
\item[]
{\bf Full SMC:} decays, showering evolution and hadronization have been
included in our simulations, using both \pythia\ and \herwig.
\end{itemize}
In both SMC setup muons (default in \pythia) and
neutral pions were assumed as stable particles. All other particles and
hadrons were allowed to be stable or to decay ac\-cor\-ding 
to the default implementation of each SMC. 
Quark and Higgs masses, as well as 
$W$, $Z$ masses and total decay widths, were tuned to the same values 
in \pythia\ and \herwig. 
On the other hand, each of the two codes was allowed to compute 
autonomously partial branching fractions in different decay channels 
for all unstable particles and hadrons. 
Multiparticle interaction effects were neglected (default in \herwig\ ).
Additionally, the intrinsic $p_T$ spreading of valence partons in incoming
hadrons in \herwig\ was assumed to be 2.5 GeV.  
Considering this setup, we found agreement between
\pythia\ and \herwig\ predictions within 5\,\%, except 
in bins where the statistics is very small.  
Beside the conceptual differences in the parton shower and
hadronization algorithms between the two SMC generators, 
written on the basis of different theoretical ideas 
($p_\perp$ vs. angular ordering, 
string model vs. cluster hadronization and preconfinement), 
a possible origin of this overall small discrepancy is the absence of the 
truncated shower in the \herwig\ prediction.
Unfortunately, however, as already mentioned, we cannot check the last point 
within the \powhegbox\ framework.  
In any case, the modest entity of the discrepancy means that the 
effect of the truncated shower, not included in our analysis, is small. 
Further explanations of small discrepancies can also be 
related to the different decay channel scheme and branching fractions
in the two SMC codes.

In our computation, we adopted the following parameters:
$\sqrt{s} = 7\tev$, \texttt{CTEQ6.6M} PDF set from \lhapdf, with a
2-loop running $\alpha_s$, 5 light flavours and 
$\Lambda_5^{\overline{\mathrm{MS}}} = 226\mev$, $m_t = 172\gev$,
$m_H = 120\gev$ , $G_F = 1.16639\cdot 10^{-5}\gev^{-2}$. The
renormalization and factorization scales were chosen equal to
$\mu_0 = m_t + m_H/2$. We decided to switch on all possible decay
channels of the Higgs boson, implemented in the SMC programs.%
\footnote{In \pythia\ there are two more decay channels 
than in \herwig, and partial decay fractions in each leptonic, bosonic
and partonic channel differ in the two codes.}
We used the last version of the SMC codes: \pythia\ 6.425 and
\herwig\ 6.520.

\begin{figure}[h!]
\begin{center}
\includegraphics[width=0.75\linewidth]{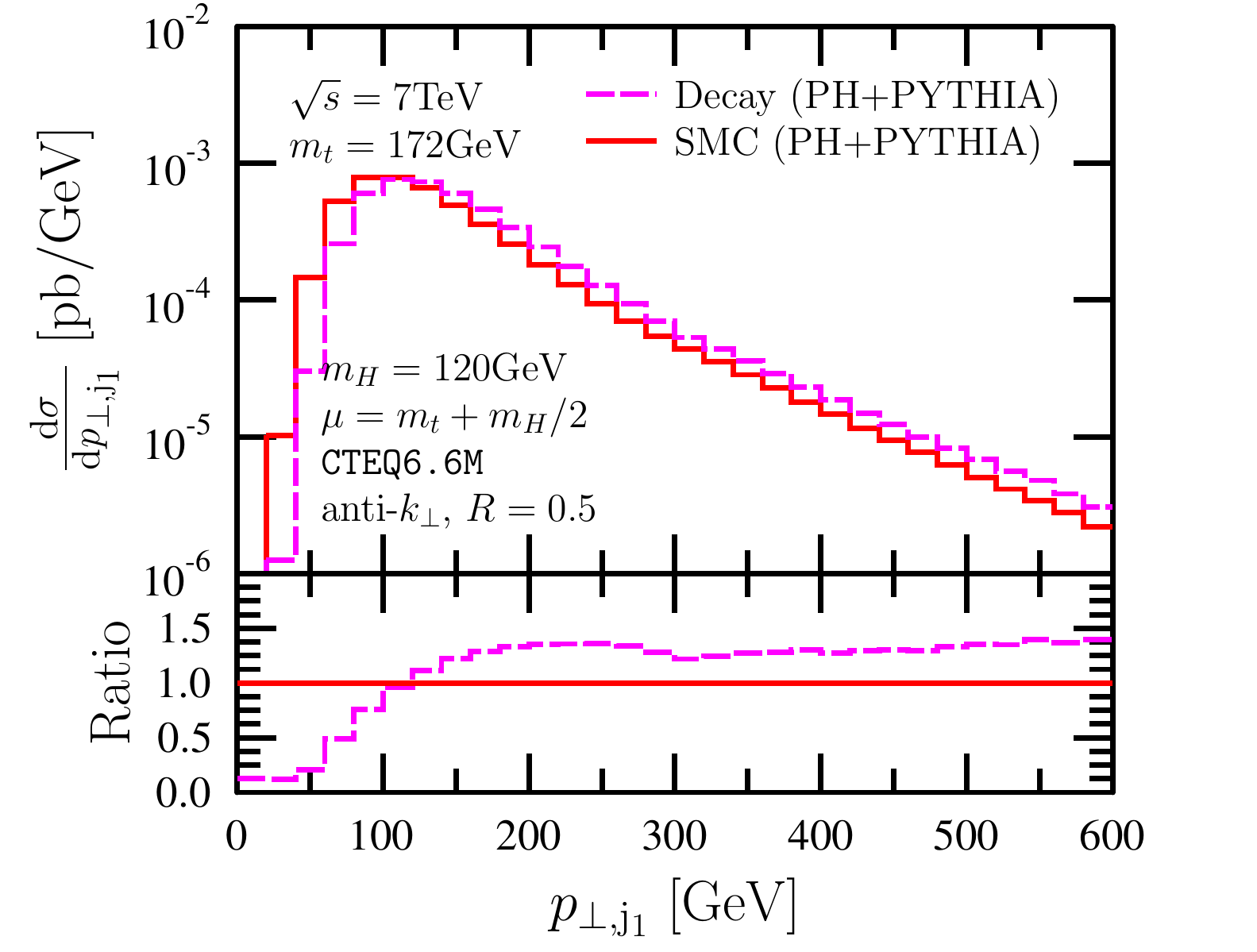}
\caption{Transverse momentum distribution of the hardest jet. 
The results at 
the decay level (dashed line) 
are compared to the ones
at the SMC level (solid line), obtained
by interfacing \powheghelac\ to \pythia\ . No selection cuts were applied.}
\label{fig:pTj}
\end{center}
\end{figure}

\begin{figure}[t!]
\begin{center}
\includegraphics[width=0.75\linewidth]{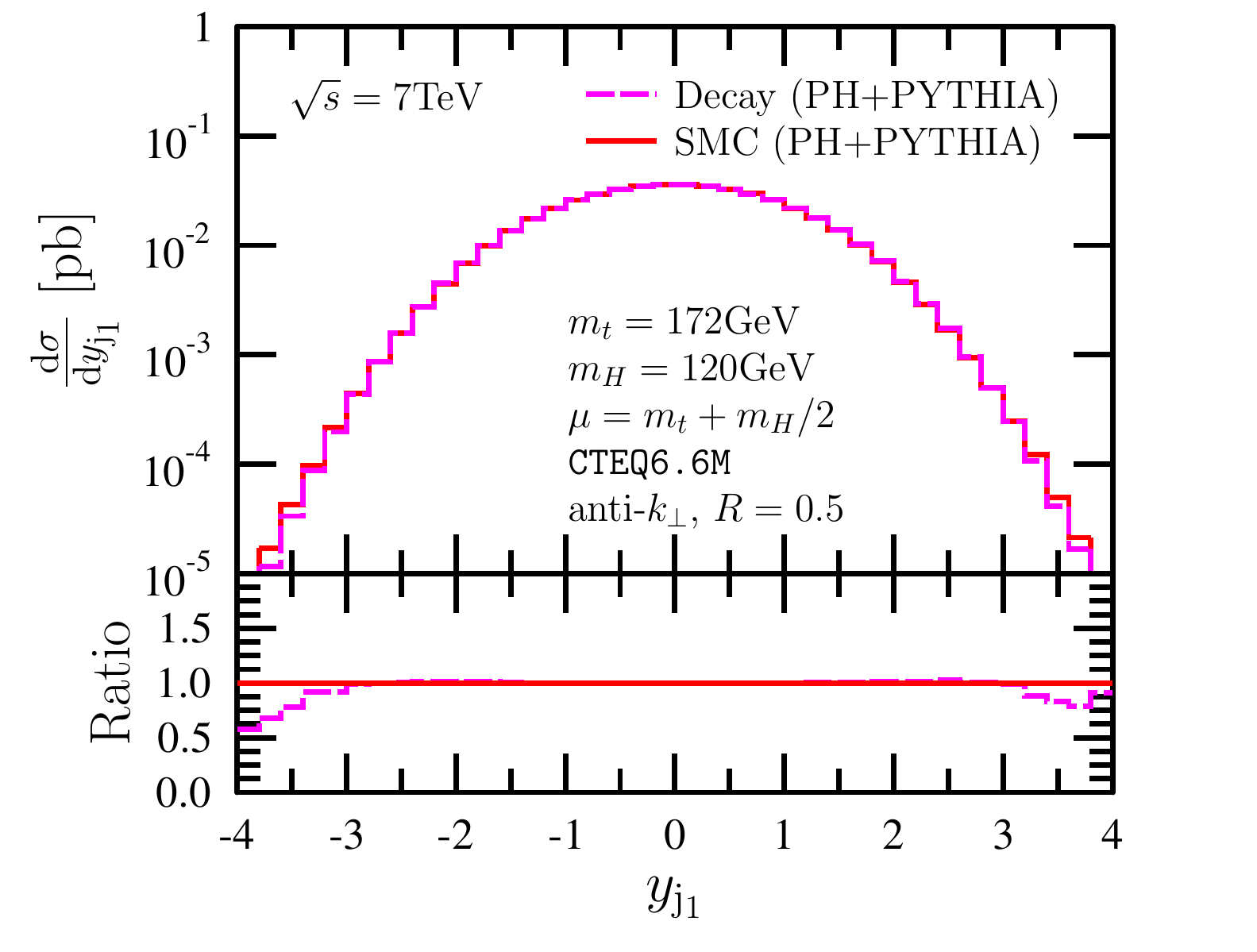}
\caption{The same as Fig.~\ref{fig:pTj}, as for the 
rapidity of the hardest jet.}
\label{fig:yj}
\end{center}
\end{figure}

We studied the effect of the full SMC by comparing distributions at the 
decay and SMC level. As the number of particles is very different at the end of the two stages, we first made such a comparison without any
selection cut, in order to avoid the introduction of any bias. This way 
the cross-section at all levels is indeed exactly 
the same. We found $\sigma_{\powheghelac} = \sigma_{\powheghelac+{\rm{DECAY}}}
=\sigma_{\powheghelac+{\rm{SMC}}}=$ 
$95.872$ $\pm$ $0.007$ fb (here and in all following $\sigma$ predictions 
the quoted uncertainties are the statistical ones only). 
As an illustrative example, we present the distributions of the
transverse momentum and rapidity of the hardest jet, \pTj\ and $y_{j_1}$,
in \fig{fig:pTj} and \fig{fig:yj}, respectively. 
The jets are reconstructed through the $\mathrm{anti}- k_\bot$
algorithm with $R=0.5$, 
by using \fastjet\ 2.4.3 \cite{Cacciari:2008gp}. 
One can observe a rather significant softening in the transverse momentum
spectrum as going from 
results at the decay level to full SMC ones. 
On the other hand, the effect of the shower on the rapidity
of the hardest jet is almost negligible and rather homogeneous.
We distinguished several classes of jets including final state 
emissions, according to their origin: 
jets that can be traced back to (i) first radiation emissions,
(ii) the decay products of the Higgs boson,
(iii) the decay products of the top and antitop quarks,
(iv) a mixing of the previous ones.  
In particular, main contributions to the \pTj spectrum 
shown in Fig.~\ref{fig:pTj} are due to jets of the (iv) 
and (iii) class.
As for the $y_{j_1}$ distribution, the tails are dominated by
jets of the (iv), (i) and (iii) classes. 

\begin{figure}[h!]
\begin{center}
\includegraphics[width=0.75\linewidth]{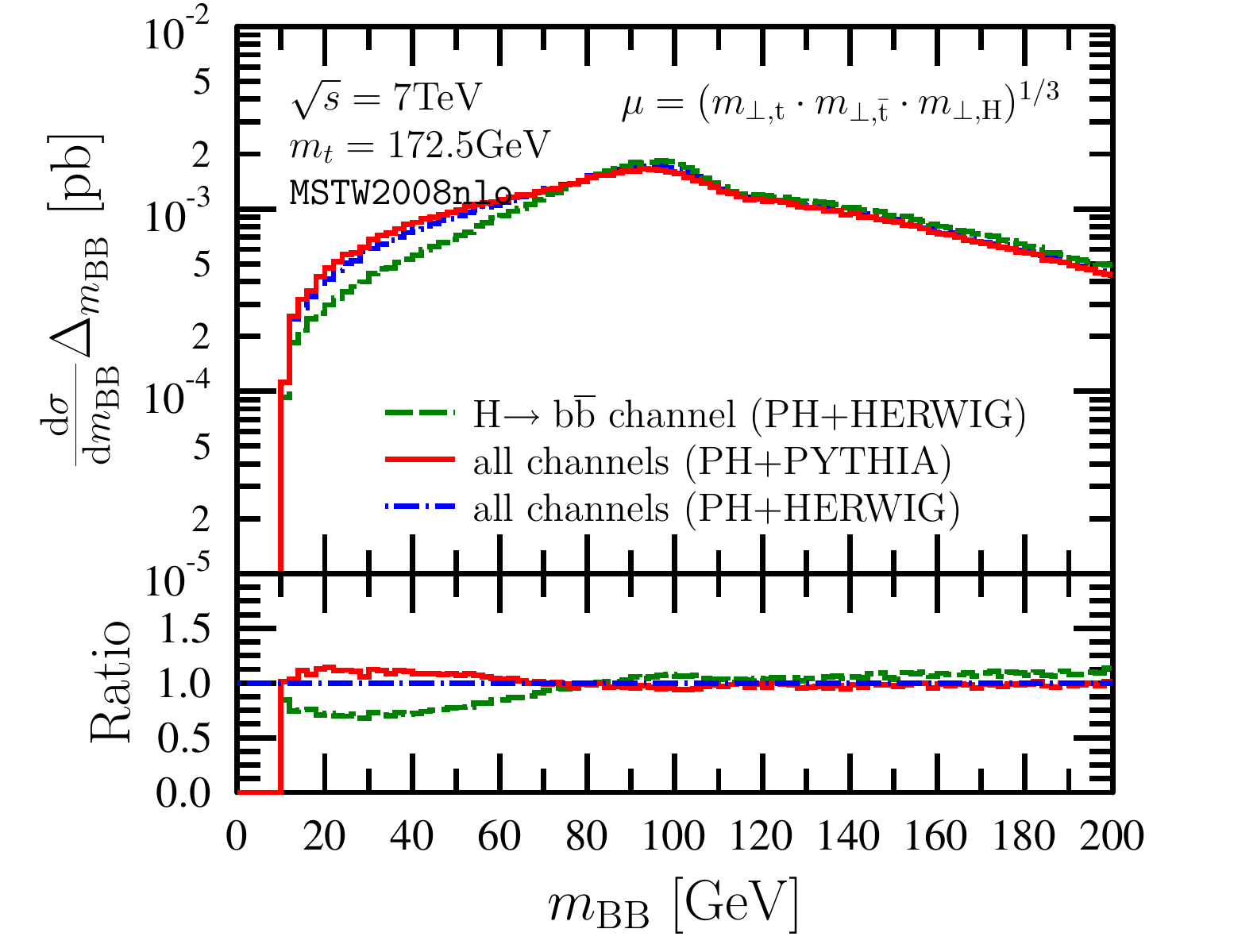}
\caption{Invariant mass distribution of the two hardest lowest-lying B hadrons at the SMC level. These results are presented in sigma per bin, just to allow an easier comparison with the results of Ref.~\cite{Hirschi:2011pa}, having been obtained in the same setup, without applying any cut. The effects of including all H decay channels, with respect to the case of a 
single H 
$\rightarrow$ b\bb\ 
channel (dashed line), were computed by interfacing
\powheghelac\ to both \pythia\ (solid line) 
and \herwig\ (dash-dotted line).}
\label{fig:mbb}
\end{center}
\end{figure}

\begin{figure}[h!]
\begin{center}
\includegraphics[width=0.75\linewidth]{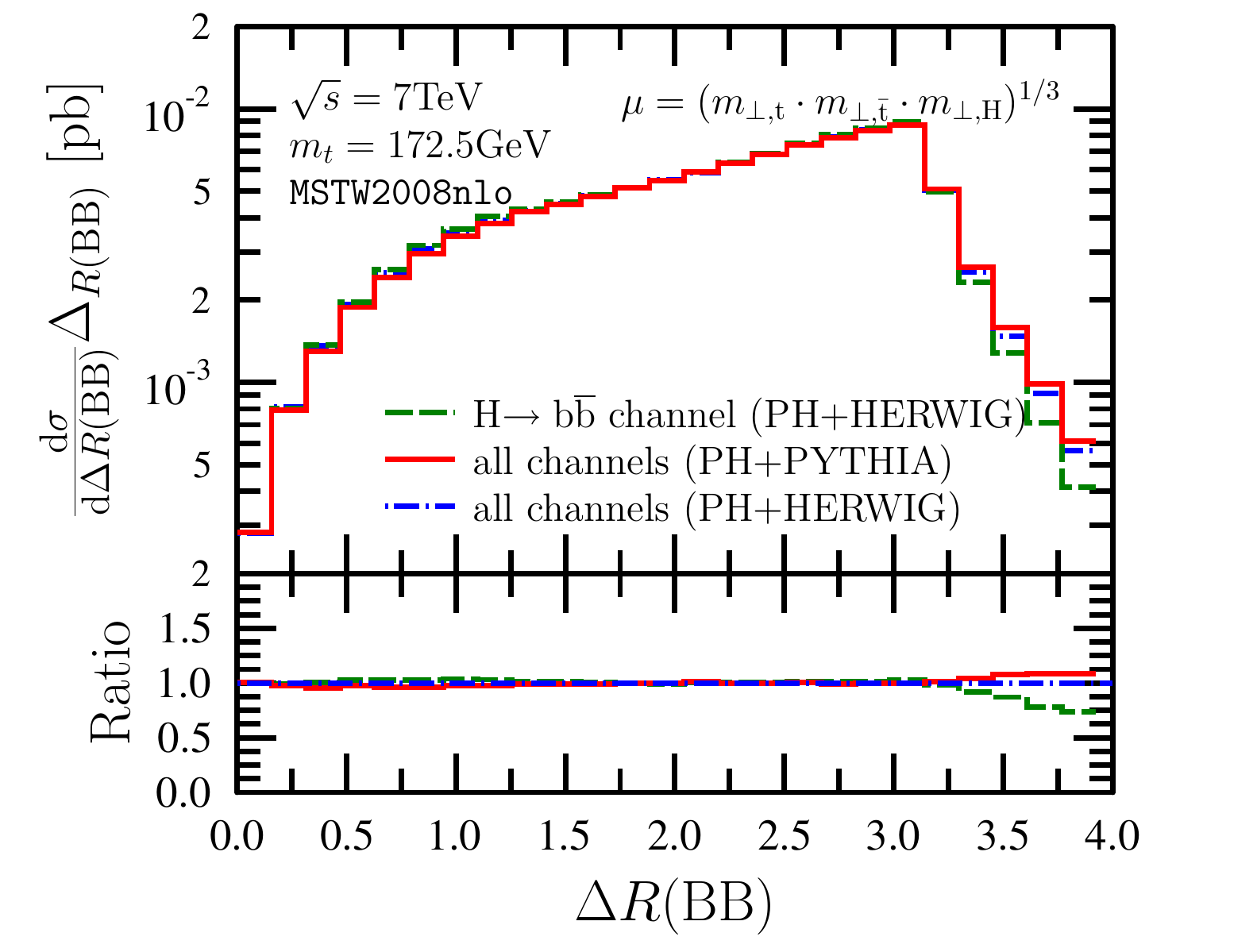}
\caption{The same as Fig.~\ref{fig:mbb}, as for the azimuth-rapidity
distance correlations of the two hardest lowest-lying B hadrons.}
\label{fig:drbb}
\end{center}
\end{figure}

In Ref.~\cite{Hirschi:2011pa} the invariant mass, $m_{BB}$, 
and the separation in the
rapidity--azimuthal-angle plane of the two hardest lowest-lying 
B-hadrons, $\Delta R_{BB}$, were studied, 
by choosing a dynamical scale for the
generation of the hard-scattering events and by taking into account 
only the H $\rightarrow$ b\bb\ decay channel. 
Besides computing all $p_T$ distributions already presented in 
that paper, 
always finding 
agreement, we also computed the aforementioned B-hadron
distributions 
reproducing the same 
simulation setup and without applying any cut, 
considering only the H $\rightarrow$ 
b\bb\ decay channel, as well as all channels. In the
former, we found agreement with the predictions of Ref.~\cite{Hirschi:2011pa}. 
The effect of the remaining channels, not studied in that work, 
produces an increase in the region 
below 80\,GeV in the $m_{BB}$ spectrum and only for large $\Delta R_{BB}$,
as shown in Figs.~\ref{fig:mbb} and~\ref{fig:drbb}. 
Pairs of B-hadrons both including quarks that can be 
traced back to Higgs decays only populate the region of the 
$m_{BB}$ spectrum below $m_H$. The region 
above $m_H$ is instead dominated by pairs with at least 
one B-hadron 
that  can be traced back to a (anti)top decay.   
 

\begin{figure}
\begin{center}
\includegraphics[width=0.75\linewidth]{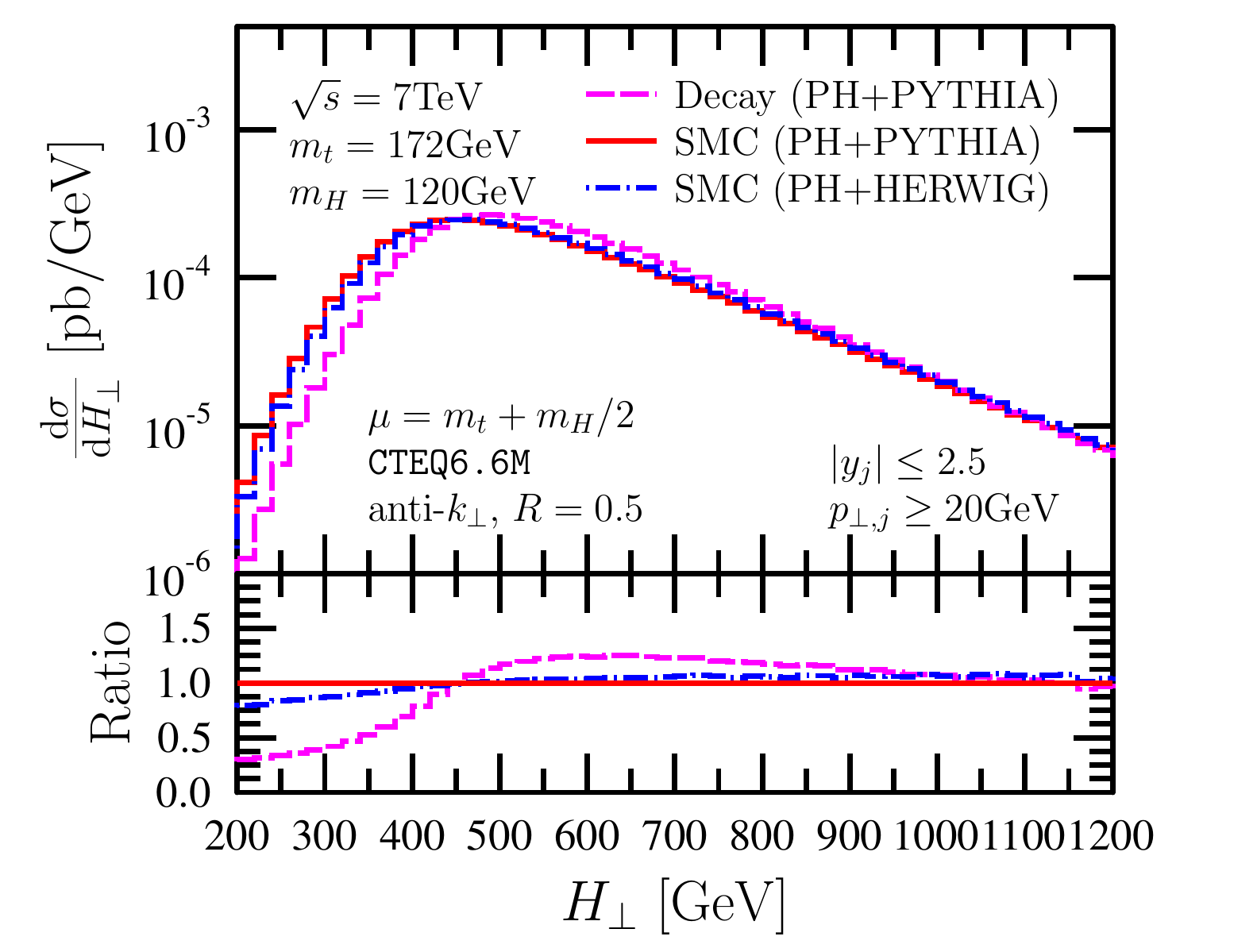}
\caption{Distribution of the sum of the transverse momenta of 
all particles, at the decay level (dashed line) 
and at the SMC level, as obtained by interfacing \powheghelac\ to \pythia\
(solid line) and \herwig\ (dash-dotted line). Only hadronic
cuts were applied.}
\label{fig:ht}
\end{center}
\end{figure}

\begin{figure}
\begin{center}
\includegraphics[width=0.75\linewidth]{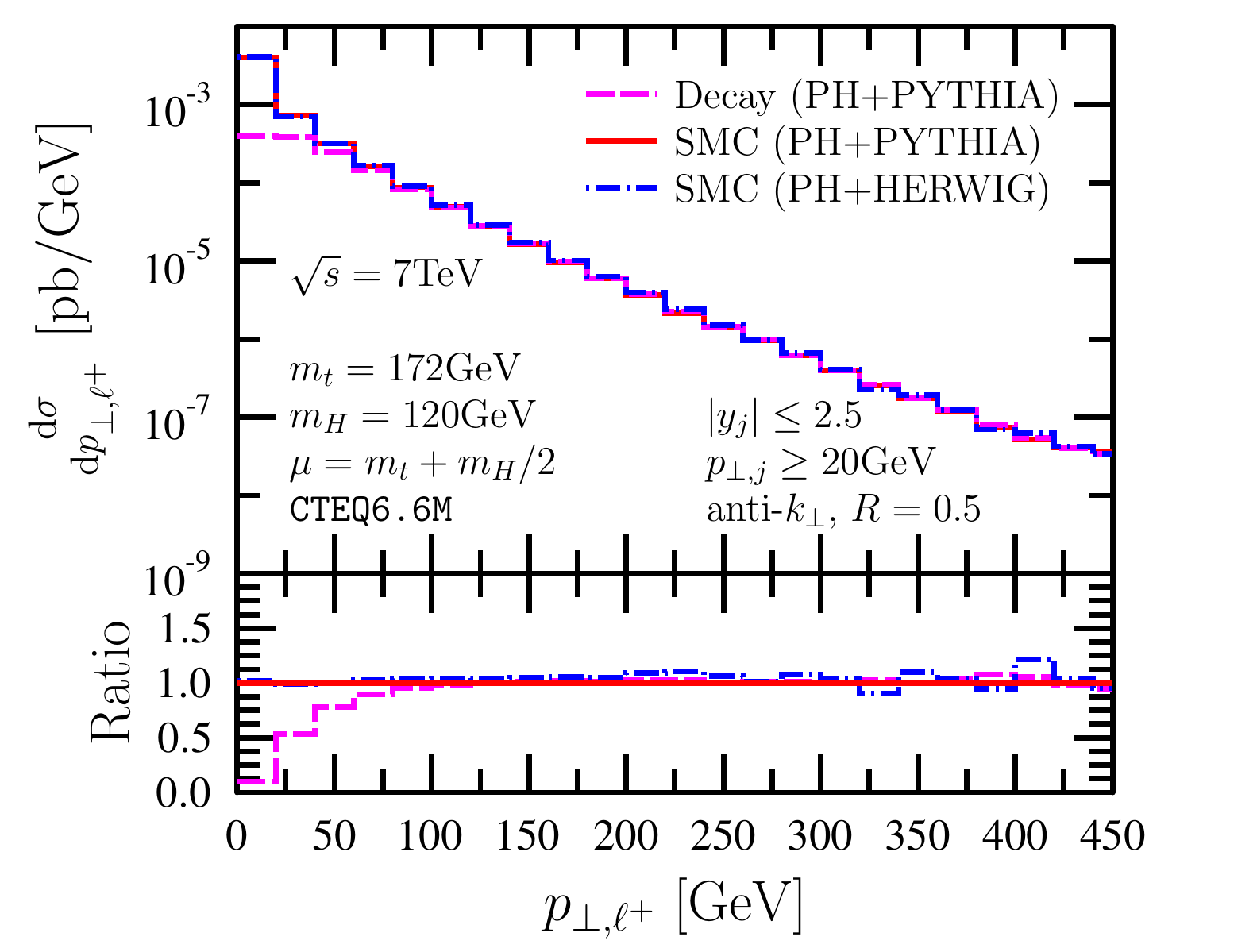}
\caption{The same as Fig.~\ref{fig:ht}, as for the 
transverse momentum of all antileptons.}
\label{fig:pTl}
\end{center}
\end{figure}

\begin{figure}
\begin{center}
\includegraphics[width=0.75\linewidth]{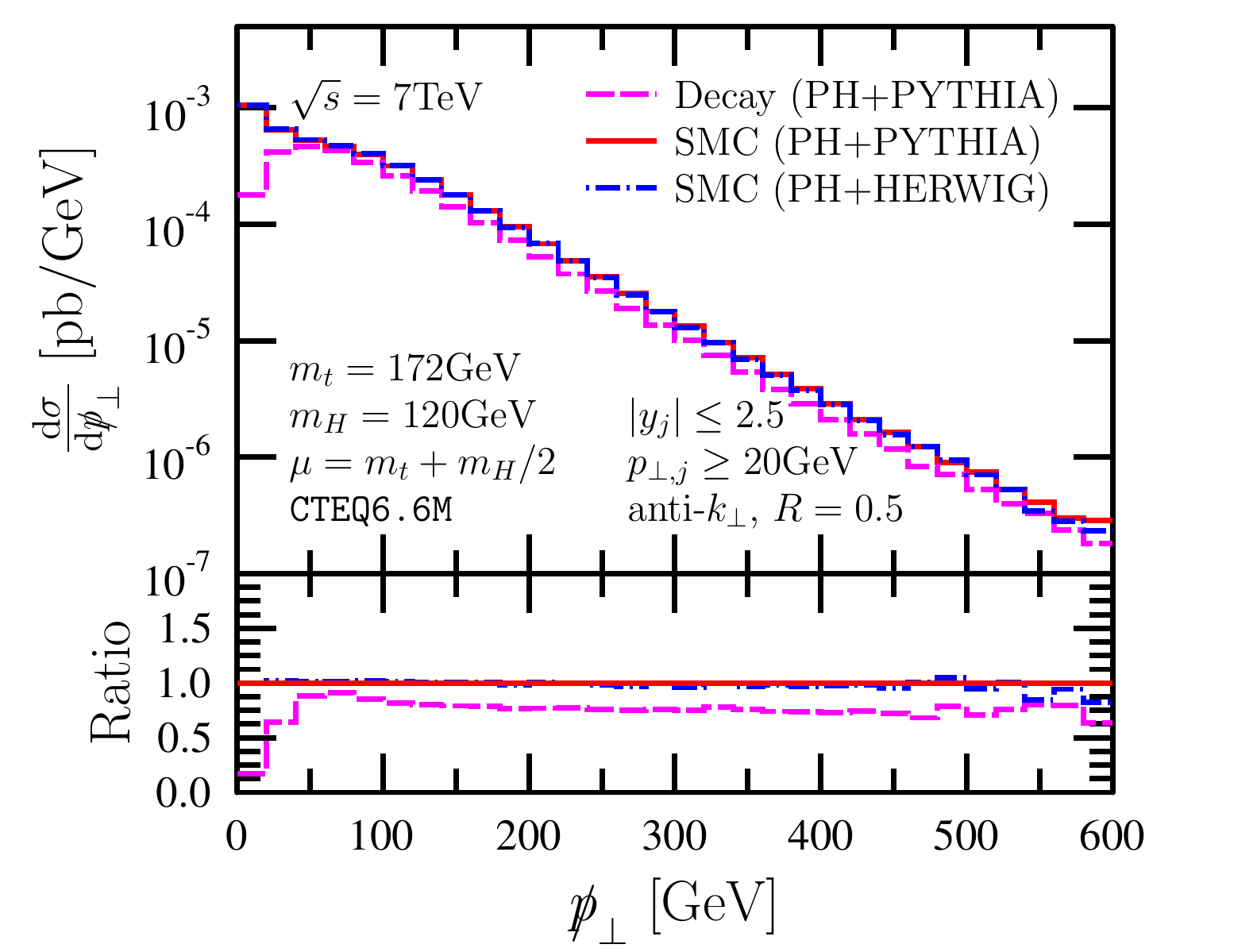}
\caption{The same as Fig.~\ref{fig:ht}, as for the total
missing transverse momentum.}
\label{fig:pTmiss}
\end{center}
\end{figure}

The effects of the decay and shower also depend 
on the selection cuts.
While the typical selection cuts include both leptons and hadrons, 
we started with a restricted set, 
not involving any leptonic cut, 
in particular as for the number of leptons. This is motivated by the fact
that this number can be quite different at the decay and
shower level, as the shower produce many secondary leptons.   
Thus, coming back to our setup,  
we considered and implemented cuts only on jet variables:
(i) $p_{\bot,\mathrm{min}}^{\mathrm{j}} = 20\gev$ and
(ii) $|y^{\mathrm{j}}|\le 2.5$ for all jets, 
(iii) $\#\mathrm{jets}\ge 4$ in each event. 
We show the distribution of the scalar sum of all transverse momenta in the
event, \HT\ in \fig{fig:ht}. We can see that 
the spectrum becomes softer due to showering effects, with respect to the one 
computed at the decay level, as can be understood 
on the basis of PS physics. The same is true and even
more evident if one singles out the hadronic component of \HT, as well
(not shown). 
On the other hand, the effects of the shower on the (anti-)lepton 
transverse momentum,
\pTl, and the missing transverse momentum, \pTmiss, 
as shown in Figs.~\ref{fig:pTl} and~\ref{fig:pTmiss}, 
are small and rather uniform, except for a significant increase for
small values, which is due to secondary leptons produced in the shower.
The cross-sections at the decay and at the SMC level, 
after the hadronic cuts listed above, amount to
$\sigma_{\powheghelac+\mathrm{DECAY}} = $  $92.29$ $\pm$  $0.01$ fb,
$\sigma_{\powheghelac+\pythia} = $  $90.46$ $\pm$  $0.01$ fb and
$\sigma_{\powheghelac+\herwig} = $  $90.99$ $\pm$  $0.01$ fb.

\section{\label{sec:predictions} Predictions}

In this section we present predictions for 
Ht\bt\ production
with parton shower and hadronization
effects at LHC. In addition to the jet cuts (i--iii) mentioned above, we also
applied selection cuts on the leptonic variables:%
\footnote{Similar cuts are applied by the LHC experiments.} 
(iv) we focused on the dileptonic channel, asking for exactly 
one $\ell^+$ and one $\ell^-$ with
(v) $p_{\bot,\mathrm{min}}^{\mathrm{\ell^\pm}} = 20\gev$ and 
(vi) $|y^{\mathrm{\ell^\pm}}| \le 2.5$, whereas the transverse missing energy
of the event was constrained to 
(vii) $\slashed{E}_{\bot,\mathrm{min}} \ge 30\gev$.

\begin{figure}
\begin{center}
\includegraphics[width=0.75\linewidth]{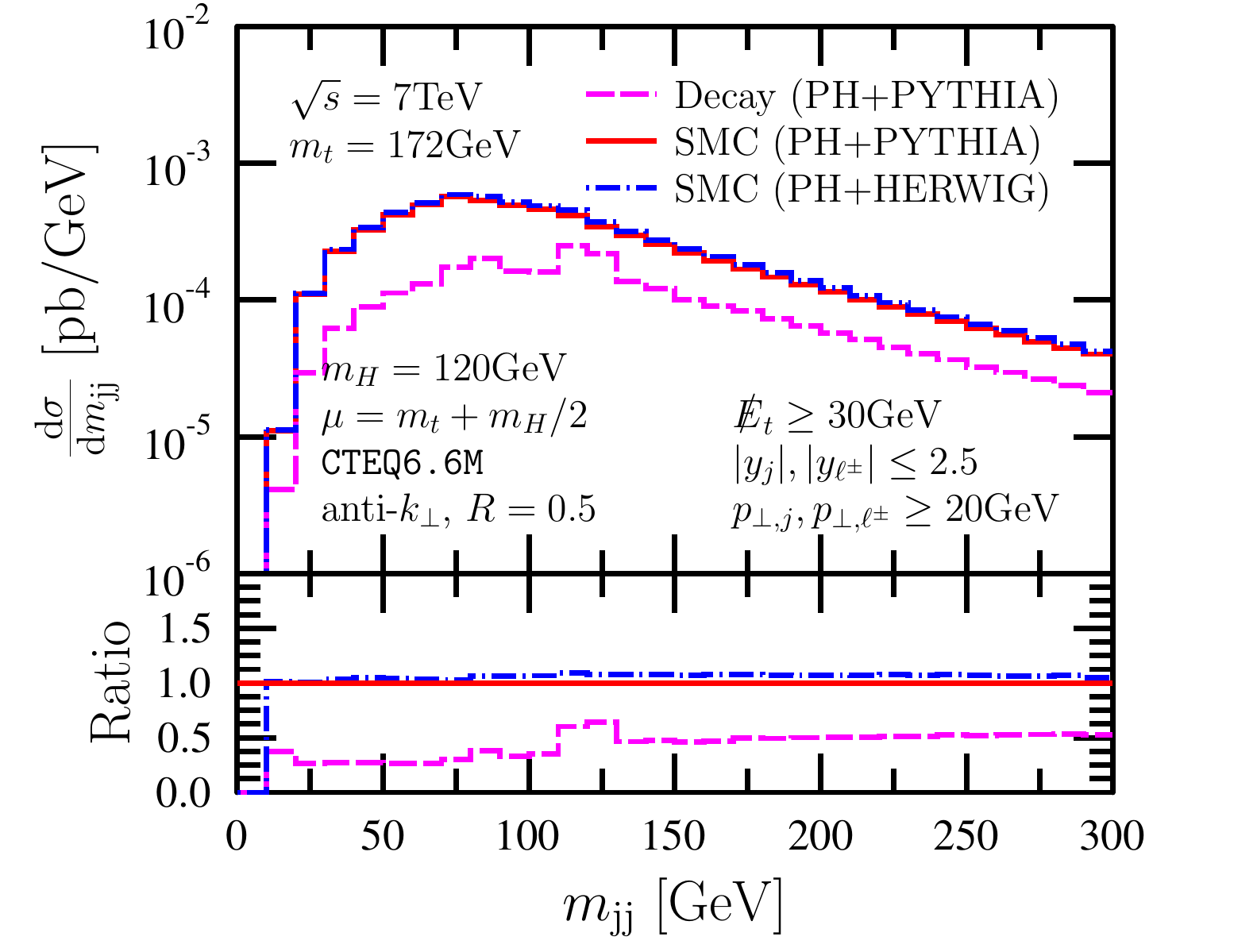}
\caption{Invariant mass distribution of all jet pairs. Results
at the decay level (dashed line) are shown together with results
after showering and hadronization, 
obtained by interfacing \powheghelac\ predictions to
both the \pythia\ (solid line) and \herwig\ (dash-dotted line) SMC.
The full set of hadronic and leptonic cuts was applied.}
\label{fig:mjj}
\end{center}
\end{figure}

In \fig{fig:mjj} we present the distribution of the invariant
mass of 
all jet pairs. Here the effect of the shower is again
quite significant. In particular, there is a small bump around
the Higgs mass, 
as already noticed in the literature~\cite{Binoth:2010ra},
visible in the data at the decay level,  
which is completely washed out when PS is included.
The same is true for the invariant mass distribution
of the two hardest jets (not shown).

The cross-section after cuts, at the SMC level, were found to be
$\sigma_{\powheghelac+\pythia} = $ $5.376$ $\pm$ $0.010$ fb
and
$\sigma_{\powheghelac+\herwig} = $ $5.521$ $\pm$ $0.011$ fb.

One of the biggest differences in the results produced by \pythia\ 
and \herwig\ interfaced to the \powhegbox\
noticed in Ref.~\cite{Oleari:2011ey} in the study of a different process was 
the observation that \herwig\ gives rise to hard jets 
more central than \pythia. 
We observe the same trend in our results, but by far to a lesser extent. 
In particular, in the bins around zero rapidity the ratio between 
the rapidity distributions
of \powheghelac\ + \herwig\ and \powheghelac\ + \pythia\ found in
our study amounts 
to maximum 1.05, both in case of the hardest and the second hardest jet. 
On the other hand, the agreement between the two SMC, as for
the rapidity distributions of leptons and antileptons, was found to
be even closer.

\section{\label{sec:conclusions} Conclusions}

We would like to conclude this letter by emphasizing that interfacing
NLO calculations, as structured in HELAC-NLO, with PS and hadronization
effects, within the POWHEG framework, open the exciting possibility of
realistic, precise and reliable simulations of pp hard-scattering 
collisions, leading to multiparticle production, up to the hadron level, 
as detected in the accelerator experiments. We have presented
predictions for the signal, pp $\to$ t\bt H production, illustrating
their consistency with existing calculations, as well as the
phenomenological potential of our approach. 
A thorough analysis, including the most important backgrounds, notably
pp $\to$ t\bt b\bb, within the same framework, is in progress and will
be reported elsewhere. The event files produced by
\powheghelac, together with further detail and results of our project,
are available at 
{\texttt{http://grid.kfki.hu/twiki/bin/view/DbTheory/TthProd}}.

\section*{Acknowledgements}
This research was supported by the HEPTOOLS EU program MRTN-CT-2006-035505,
the LHCPhenoNet network PITN-GA-2010-264564,
the Swiss National Science Foundation Joint Research Project SCOPES
IZ73Z0\_1/28079, the MEC project FPA 2008-02984 (FALCON),
and the T\'AMOP 4.2.1./B-09/1/KONV-2010-0007 project.
We are grateful to M.~Kr\"amer for providing us the data files with their
NLO predictions \cite{Beenakker:2002nc}.
M.V.G. is grateful to the Departamento de F\'{i}sica Te\'orica y del Cosmos 
of Granada University for hospitality.
A.K. is grateful to G.~Bevilacqua for useful discussions on 
the original \helac\ programs.

\end{document}